\documentclass[12pt]{article}

\usepackage{epsfig}
\usepackage{lscape}

\newcommand{\eq}{\begin{equation}}
\newcommand{\eqx}{\end{equation}}
\newcommand{\eqn}{\begin{eqnarray}}
\newcommand{\eqnx}{\end{eqnarray}}

\begin{document}

\title{\Large\bf
\begin{flushright}
\normalsize\rm TPJU-6/2003
\end{flushright}
\begin{flushright}
\normalsize\rm MPP-2003-72
\end{flushright}
\vskip 20 pt {\Large Recent progress in supersymmetric
 Yang-Mills quantum mechanics\\ in various dimensions}
\footnote{Presented at the Workshop on Random
Geometry, Kraków, Poland, May 15--17, 2003.} }
\author{\large  J. Wosiek\\  \\
  %
  $M.\; Smoluchowski\; Institute\; of\; Physics$\\
$Jagellonian\; University $\\
$Reymonta\; 4\;, 30-059\; Krakow,\; Poland\;$\\
}

\maketitle

\begin{abstract}
We review the last year progress in understanding supersymmetric SU(2) Yang-Mills
quantum mechanics in the $D=4$ and $10$ space-time dimensions.
The four dimensional system is now well under control and the precise spectrum
is obtained in all channels. In $D=10$ some new results are also available.
\end{abstract}
\vskip 20 pt
PACS: 11.10.Kk, 04.60.Kz\newline {\em Keywords}:
 quantum mechanics, Yang-Mills, supersymmetry\newline
\newpage

\section{Introduction}

Supersymmetric Yang-Mills quantum mechanics (SYMQM) shares many
properties of advanced field theories thereby it provides a useful
laboratory to study explicitly some of their properties
\cite{CH,HS}. It emerges from the dimensional reduction (in space)
of the full supersymmetric Yang-Mills field theory defined in the
$D$ dimensional space-time. Depending on $D$ it covers a wide
range of interesting phenomena. $D=2$ system is exactly soluble
with manifestly supersymmetric continuous spectrum and
analytically calculable Witten index
 \cite{CH,CW}.
For $D=4$ the model is nontrivial and posesses both localized and
non localized eigenstates, which have been constructed only
recently in all fermionic sectors. On the other hand in the zero
fermion sector it reduces to the zero volume limit of the pure
Yang-Mills theory well studied in the context of  lattice field
theory \cite{L,LM,VABA}. Finally the $D=10$, $SU(\infty)$ system,
with its threshold bound state and the continuum of scattering
states, is considered as a model of M-theory \cite{BFSS} and has
attracted a lot of interest (for recent review see e.g.\cite{TA}).

In this talk I would like to review the current status of the programme which
 attempts to solve these models in various dimensions
and for various gauge groups \cite{JW}. The four dimensional ($D=4$) system with the
 SU(2) gauge group,
where the main progress has been achieved,
will be discussed in the next Section. In Section 3,  a new approach to the $D=10$, SU(2)
model,
together with some preliminary results for the purely bosonic sector will be presented.

\section{D=4 Supersymmetric Yang-Mills  quantum mechanics}

Reduction of the D dimensional supersymmetric SU(2) Yang-Mills
field theory to a single point in the $d=D-1$ dimensional space
leads to the quantum mechanical system which for $D=4$ is
described by nine bosonic coordinates $ x^i_a(t) $, $i=1,2,3;
a=1,2,3$ and six independent fermionic coordinates contained in
the Majorana spinor $\psi_a^{\alpha}(t)$, $\alpha=1,...,4$.
Equivalently (in D=4) one could impose the Weyl condition and work
with Weyl spinors. Hamiltonian reads \cite{HS}
 \eqn
\label{HD4}
H=H_B+H_F,  \nonumber\\
 H_B={1\over 2} p_a^ip_a^i+{g^2\over 4}\epsilon_{abc}
\epsilon_{ade}x_b^i x_c^j x_d^i x_e^j,\\
 H_F={i g \over 2} \epsilon_{abc}\psi_a^T\Gamma^k\psi_b x_c^k,\nonumber
 \eqnx
 where
$\psi^T$ is the transpose of the real Majorana spinor, and
$\Gamma$ in D=4 are just the standard Dirac $\alpha$ matrices.

After the reduction of the three dimensional space to a single
point, the rotational symmetry of the original theory becomes the internal spin(3)
symmetry. It is generated by the angular momentum
\eq
J^i=\epsilon^{ijk}\left( x^j_a p^k_a-{1\over
4}\psi^T_a\Sigma^{jk}\psi_a\right),\label{JD4}
\eqx
with
\eq
\Sigma^{jk}=-{i\over 4}[\Gamma^j,\Gamma^k].
\eqx
The model has also a gauge invariance with the generators
\eq
G_a=\epsilon_{abc}\left(x_b^k p_c^k-{i\over 2}\psi^T_b\psi_c
\right), \label{GG}
\eqx
and is invariant under the supersymmetry
transformations generated by
\eq
Q_{\alpha}=(\Gamma^k\psi_a)_{\alpha}p^k_a + i g
\epsilon_{abc}(\Sigma^{jk}\psi_a)_{\alpha}x^j_b x^k_c. \label{QD4}
\eqx
The bosonic potential (written in the vector notation in
the color space)
 \eq
V={g^2\over
4}\Sigma_{jk}(\vec{x}^j\times\vec{x}^k)^2, \label{v}
\eqx
exhibits the famous flat directions responsible for a rich structure of the
spectrum.

\subsection{Original approach}
We shall now review a simple way to  calculate automatically
algebraic expressions for matrix elements of a wide class of
quantum observables and to construct numerically the complete
spectrum of a given system \cite{JW}. The method implements
literarily in the computer the rules which govern  the quantum
world.\newpage

\noindent{\em Quantum mechanics inside a PC}

For a moment consider a single bosonic degree of freedom. Generalization to fermions
and to more variables readily follows. Action of any polynomial observable
can be easily implemented in an algebraic program if we use the discrete eigenbasis
of the occupation number operator $a^{\dagger} a$
\eq
\{ |n> \}, \;\;\; |n>={1\over\sqrt{n!}}(a^\dagger)^n |0>. \label{basis}
\eqx
 Since the bosonic coordinate and momentum operators are
\eq
   x={1\over\sqrt{2}}(a+a^{\dagger}),\;\;\; p={1\over i \sqrt{2}}(a-a^{\dagger}),
\label{XP}
\eqx
a typical quantum observable can be
represented as the multiple actions of the basic creation and annihilation
operators\footnote{The method can be also extended
to non polynomial potentials.}.

A quantum state is a superposition of arbitrary number, $n_s$, of elementary states $|n>$
\eq
|st>=\Sigma_I^{n_s} a_I |n^{(I)}>, \label{stQ}
\eqx
and will be represented as a Mathematica list
\eq
st=\{n_s,\{a_1,\dots,a_{n_s}\},\{n^{(1)}\},\{n^{(2)}\},\dots,\{n^{(n_s)}\}\}, \label{stM}
\eqx
with $n_s+2$ elements. The first element specifies the number of elementary states
 entering the linear combination,
Eq.(\ref{stQ}). The second element is the list itself and contains all complex amplitudes
$a_I, I=1,\dots,n_s $.
Remaining $n_s$ sublists give the occupation numbers of elementary, basis states.
According to this convention an elementary state $|n>$ is represented by
$
 \{1,\{1\},\{n\} \}.
$

Next, we implement basic operations defined on states: addition,
multiplication by a number and the scalar product. They are simply programmed
as definite operations on Mathematica lists transforming them in accord
with the principles of quantum mechanics.
Creation and annihilation operators are then defined as a list-valued
functions on above lists.  According to Eq.(\ref{XP}) the action of the position
and momentum operators becomes also defined.
Then we define any quantum observable:
hamiltonian, angular momentum, generators of gauge transformations,
supersymmetry generators, etc.

Further procedure is now clear: given a particular system, define
the list corresponding to the empty state, then generate a finite
basis of $N_{cut}$ vectors and calculate matrix representations of
the hamiltonian and other operators using above rules. Next, the
complete spectrum and its various symmetry properties is obtained
by the numerical diagonalization.

By studying the cutoff dependence of the spectrum one can estimate the
systematic errors induced by restricting the Hilbert space. In many systems studied so far
convergent results were obtained before the basis grew too large.

\noindent{$Creation\; and\; annihilation\; operators\; for\; SYMQM$}\newline
Supersymmetric hamiltonian (\ref{HD4}) is polynomial in momenta and
coordinates, hence above idea can be readily applied.
To this end rewrite bosonic and fermionic variables in
terms of the creation and annihilation operators of simple,
normalized harmonic oscillators
\eq
[a_a^i,a_b^{k\dagger}]=\delta^{ik}_{ab},
\;\;\;\{f_a^{\rho},f_b^{\sigma\dagger}\}=\delta_{ab}^{\rho\sigma},
\;\;\rho, \sigma = 1,2 , \label{aaff}
\eqx
such that the canonical (anti)commutation relations
\eq
[x_a^i,p_b^k]=i\delta^{ik}\delta_{ab} ,\;\;
\{\psi_a^{\alpha},\psi_b^{\beta}\}=\delta_{ab}^{\alpha\beta}.
\label{acom}
\eqx
are preserved.
Standard extensions of Eq.(\ref{XP}) for bosonic variables read
\eq
   x^i_a={1\over\sqrt{2}}(a^i_a+a_a^{i \dagger}), \;\;\; p^i_a={1\over i
\sqrt{2}}(a^i_a-a_a^{i \dagger}).  \label{XPD4}
\eqx
For fermions, the following representation
for a quantum hermitean Majorana spinor was used
\eq
\psi_a={1+i\over
2\sqrt{2}} \left( \begin{array}{c}
                            -   f_a^{1} - i f_a^{2} + i f_a^{1\dagger} +   f_a^{2\dagger}
\\
                            + i f_a^{1} -   f_a^{2} -   f_a^{1\dagger} + i f_a^{2\dagger}
\\
                            -   f_a^{1} + i f_a^{2} + i f_a^{1\dagger} -   f_a^{2\dagger}
\\
                            -i  f_a^{1} -   f_a^{2} +   f_a^{1\dagger} + i f_a^{2\dagger}
\\
                                    \end{array} \right).
\eqx

\noindent{$ The\; basis\; and\; the\; cutoff$}\newline
 The complete Hilbert space is spanned by all
independent polynomials of creation operators ${a_b^i}^{\dagger}$
and ${f_c^\sigma}^{\dagger}$ acting on the empty state
\eq
|(0,0,0),(0,0,0),(0,0,0),(0,0,0),(0,0,0)> ,\label{D4vac}
\eqx
which in the Mathematica "representation" reads
\eq
\{1,\{1\},\{\{0,0,0\},\{0,0,0\},\{0,0,0\},\{0,0,0\},\{0,0,0\}\}\}.
\eqx
By construction, the first three vectors (in color) specify bosonic, and the
last two fermionic, occupation numbers. In practical applications
we shall work in the restricted Hilbert space containing at most
$B$ bosonic quanta in total. Hence the gauge and rotationally invariant
cutoff $N_{cut}$ is defined as
\eq
\Sigma_{i,b} a_b^i {a_b^i}^{\dagger} \equiv B \le B_{max}\equiv N_{cut}.
\eqx
Since the Pauli principle admits only six
Majorana fermions in this system, there is no need to restrict the fermion number $F$.

Local gauge invariance is taken into account by constructing only the physical, i.e.
gauge invariant basis.
To create all independent, gauge invariant
 states at fixed $F$ and $B$ consider all possible contractions of color indices in
a creator of $(F,B)$ order
\eq
a^{i_{1}\dagger}_{a_{1}}...a^{i_{B}\dagger}_{a_{B}}f^{\sigma_{1}\dagger
}_{b_{1}}..f^{\sigma_{F}\dagger}_{b_{F}},
\eqx
 for all
values of the spatial indices $i$ and $\sigma$. All color contractions fall naturally into
different
{\em gauge invariant classes}. Creators from different classes differ by color
contractions {\em between}
 bosonic and fermionic operators. For example
\eq
a^{i\dagger}_a a^{j\dagger}_a a^{k\dagger}_b a^{l\dagger}_b f^{\sigma\dagger}_c
f^{\rho\dagger}_c  \label{C1}
\eqx
and
\eq
a^{i\dagger}_a a^{j\dagger}_a a^{k\dagger}_b a^{l\dagger}_c f^{\sigma\dagger}_c
f^{\rho\dagger}_b
\eqx
are in different gauge invariant classes.
Creators of odd order are constructed with one triple contraction. For example
\eq
\epsilon_{cde}a^{i\dagger}_a a^{j\dagger}_a a^{k\dagger}_b a^{l\dagger}_b a^{m\dagger}_c
f^{\sigma\dagger}_d f^{\rho\dagger}_e
\eqx
and
\eq
\epsilon_{cde}a^{i\dagger}_a a^{j\dagger}_a a^{k\dagger}_b a^{l\dagger}_c a^{m\dagger}_d
                                     f^{\sigma\dagger}_e f^{\rho\dagger}_b   \label{C2}
\eqx also belong to different gauge invariant classes. To select
linearly independent states we used again the rules of "quantum
algebra". Therefore, at fixed $F$ and $B$ the final procedure is
as follows: $(i)$ identify all gauge invariant classes of creators,
$(ii)$ loop over all values of spatial indices and for each
$i_1,...,i_B,\sigma_1,...,\sigma_F$ create corresponding state
from the empty state, Eq.(\ref{D4vac}), $(iii)$ identify  and reject
linearly dependent vectors, $(iv)$ orthonormalize the remaining set of
states.

Given the basis it was then a simple matter to calculate automatically
matrix representation of the hamiltonian and other observables. \vspace*{.3cm}

\noindent $Early\; results$

\begin{figure}[htb]
\epsfig{width=12cm,file=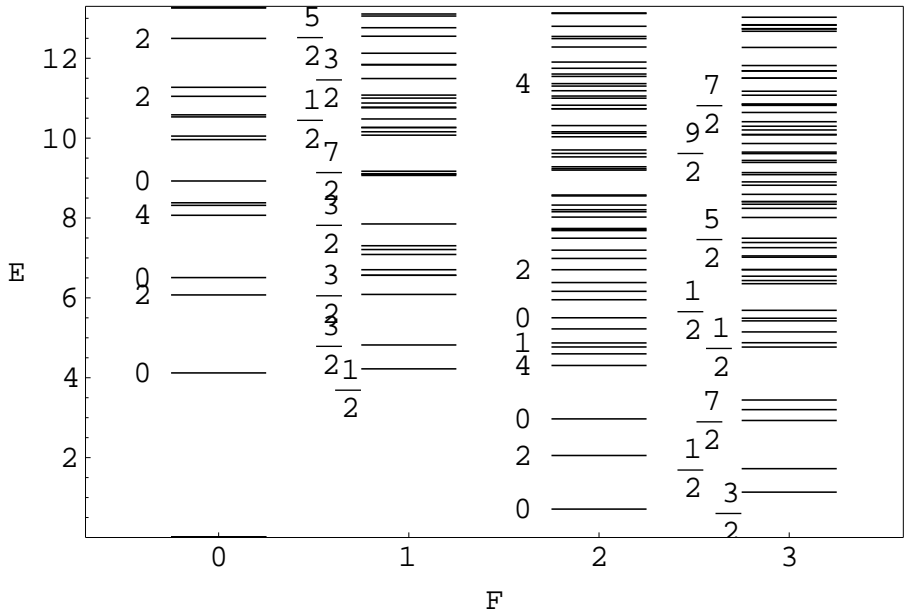}
\caption{ Spectrum of D=4 supersymmetric Yang-Mills quantum
mechanics.} \label{f1}
\end{figure}

In Fig. 1  we quote the spectrum obtained in Ref.\cite{JW} in all
four independent fermionic sectors\footnote{The hamiltonian has also the particle-hole
symmetry.}.
The cutoff in this calculation
was varying between $B_{max}=8$ for $F=0$ and $B_{max}=5$ for the
largest (i.e. most difficult) $F=3$ sector. The angular momentum
of a sample of states is also shown. It was deduced from the
degeneracy of the SO(3) multiplets and independently from the explicit
representation of the angular momentum in our basis.

\begin{figure}[htb]
\epsfig{width=12cm,file=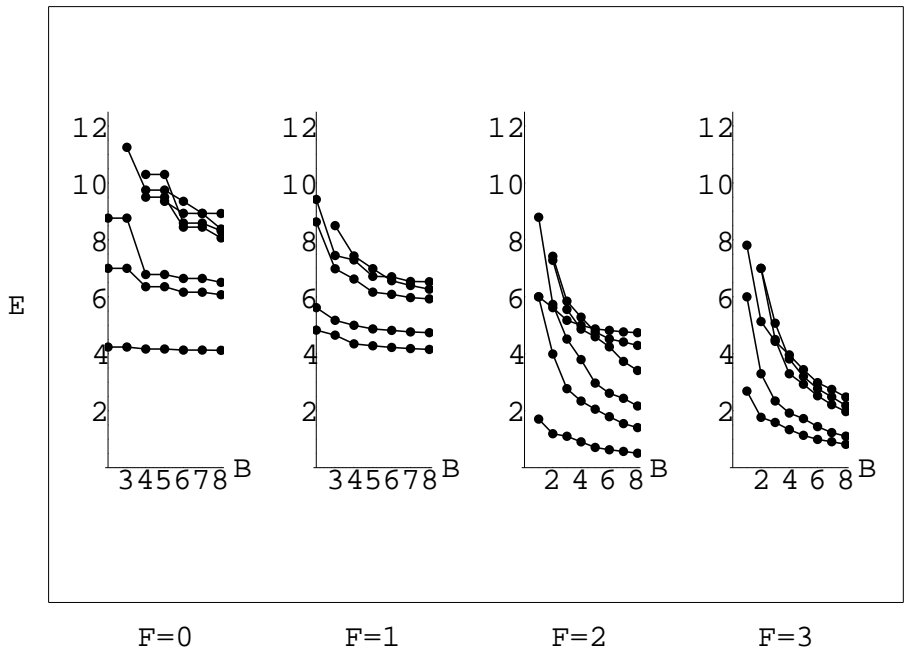}
\caption{ Cutoff dependence of the first few eigenenergies of the
hamiltonian (\ref{HD4}).} \label{f2}
\end{figure}

Later on this results were upgraded, with the considerable
numerical effort, to $B_{max}=8$ in all sectors \cite{KW}. Fig. 2
is the outcome of the latter calculations which took a couple of
months of a 600 MHz ALPHA workstation. The cutoff dependence
displayed in this Figure confirmed and quantified the general
expectation that the system has both continuous and discrete
spectrum. Discrete, localized states (which manifest themselves
here as quickly convergent levels) appear for $F=0$ and $F=1$
while the non localized states, with substantially slower
convergence, are seen in the "fermion rich" sectors with $F=2$ and
$3$. This was expected from
 the flat valley nature of the potential, Eq.(\ref{v}), combined with the supersymmetry
\cite{LNDW}.
A series of other results about the supersymmetric structure of the system, Witten index,
etc.
was also obtained.

Further progress was achieved recently by exploiting analytically the symmetries
of the system.

\subsection{New developments}

\noindent $Separation\; of\; variables$\newline Above results have
been beautifully confirmed and extended with the aid of the
"almost analytical" approach pioneered by Savvidy \cite{SAV} and
developed by van Baal in slightly different context \cite{vBN}.
Decomposing solution of the nine dimensional Schrodinger equation,
in the $F=0, J=0$ and $F=2, J=0$ channels into covariant tensors,
he reduced the problem to a numerically tractable set of coupled
ordinary differential equations. When adapted to our case his
method can push the cutoff as high as $B_{max}=39$ in these two
channels, see Fig.3. Now the discrete, localized, and quickly
convergent with the cutoff states with $F=0$ are clearly seen,
their energy determined with a very high precision. Moreover, the
intricate  nature of the solutions with two fermions is also
evident. The flat lines signal again the localized bound states of
two gluinos, while slowly falling with the cutoff levels
correspond to the non localized states from the continuous
spectrum. Some of the bound states in $F=0$ and $F=2$ sectors have
degenerate energies as required by supersymmetry. Other
supersymmetric partners must then be located in fully fermionic
 ($F=1,3)$ sectors for which the method has not been yet generalized.

\begin{figure}[htb]
\epsfig{width=14cm,file=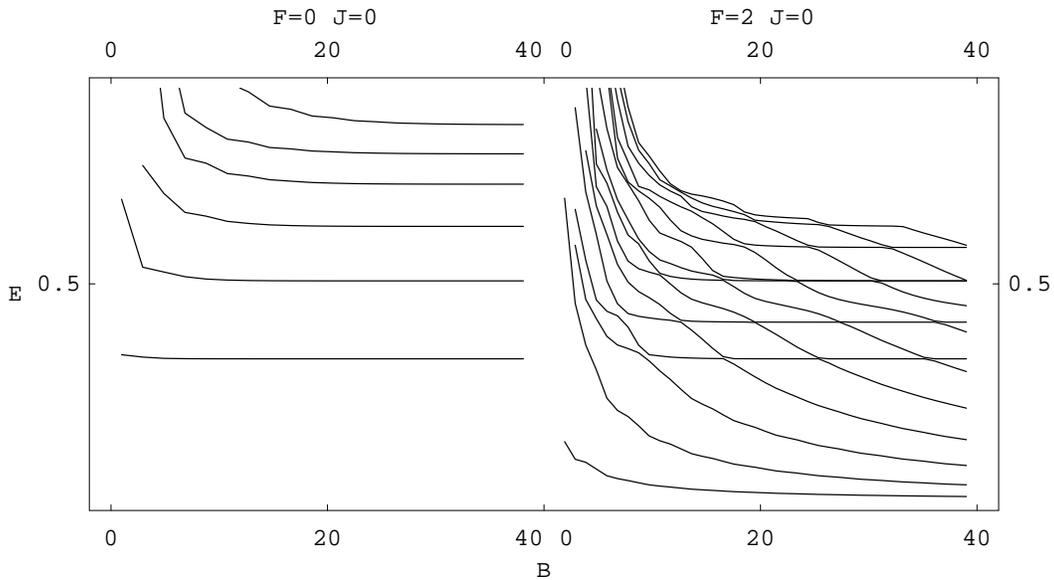}
\caption{ High cutoff results from van Baal approach.}
\label{f3}
\end{figure}

\noindent $High\; cutoff\; solutions\; in\; all\;
channels$\newline Similar high precision results are now available
for all angular momenta and in all fermionic sectors \cite{CW1}.
To this end we have extended the recursive approach first applied
to the $D=2$ system \cite{CW}. In the two dimensional case
recursions, which relate the matrix elements in a bigger basis
with those in the smaller basis, were simple enough to be solved
analytically. For the $D=4$ this is not the case any more. However
they speed up dramatically computations allowing to reach high
precision spectra in all fermionic sectors and all angular
momentum channels. Second, rotational invariance is now fully
employed, by projecting into channels of given angular momentum.
Consequently there is no need to diagonalize very large matrices.
Both these improvements result not only in the more precise
spectra, but in much more satisfactory understanding of the
supersymmetric structure of the $D=4$ system, including complete
classification of all supermultiplets.

\noindent $Analytical\; study\; of\; continuous\; spectra\; in\;
cut\; systems$\newline Finally, in a recent development a new
scaling in the continuous spectra of a cut quantum systems was
found \cite{TW}. This provides an important tool for studying the
continuum limit of non localized states and their interactions.

To summarize this section, the technology to study quantitatively $D=4$, SU(2), SYMOM is
now
available and one can readily attack more subtle problems like scattering \cite{BERS} or
thermodynamics
of this system \cite{KAB}.

We now turn to, much more difficult, $D=10$ model whose
quantitative study with presently discussed methods is just
beginning (see \cite{HS, UPP} for other attempts).

\section{TWS scheme}
In higher dimensions only the original approach is available and the spectrum of lowest
zero-volume glueballs was obtained for all $4\le D \le 10$ in the $F=0$ sector\cite{JW2}.
The number of states
grows rapidly with D and consequently at $D=10$ only $B \le 4$ were reached in practice.
In fermionic sectors situation is yet worse.
The difficulties are two-fold: a) bases become too large to store, and b) time required
for calculating the full hamiltonian matrix grows prohibitively large. The TWS scheme,
presented below, offers substantial reduction of both parameters.

Consider the (B,F) sector of the Hilbert space, i.e. the one with
B bosons and F fermions. The basis vectors before
orthonormalization can be uniquely characterized by the compound
index
\eq
I=\{C,i^1,\dots,i^B,\sigma^1,\dots,\sigma^F\}, \label{ix}
\eqx
since given $I$ the actual state can be readily created by
the compound creator from the gauge invariant class $C$ (c.f.
\ref{C1}-\ref{C2}). Many properties of the mixing matrix and of
the hamiltonian matrix can be deduced from the index $I$ alone
without need to create the actual state. Therefore we shall
abandon where possible the  lengthy Mathematica representation of
states and use instead much more compact indices (\ref{ix}) . This
virtually eliminates the first problem. Second difficulty is also
reduced by observing that for polynomial hamiltonians most of the
matrix elements are zero and many of non-zero ones are actually
the same. We shall therefore expose the sparse nature of the
hamiltonian matrix and identify equal matrix elements. This can be
done using indices (\ref{ix}) only, and will lead to the new
classification of basis states.

We begin with the properties of the mixing matrix $ M_{IK}=<I|K>,
$ also referred to as the norm matrix.
\subsection{Norm matrix and new classification of states}
For the sake of simplicity we restrict the analysis to the $F=0$ sector.
Generalization for fermions is possible. Consider the generic element of the mixing matrix
\eq
M_{IK}=<I|K>=<C_I,i^1,\dots,i^{B_I}|C_K,k^1,\dots,k^{B_K}>.
\eqx
Obviously only states which contain the same quanta can mix, therefore
\eq
M_{IK}\ne 0\;\; if\;\; B_I=B_K= B, \{i^1,\dots,i^{B}\} \sim \{k^1,\dots,k^{B}\},
\label{mix}
\eqx
where the "$\sim$" means that the values of all \{ i \} and \{ k \} indices coincide
up to a permutation. On the other hand states from different gauge invariant classes
can mix in general.

\noindent $ Words$ \newline This simple observation suggests to
group all states with given $B$ into $words$. A word $W$ is
defined as a set of all states which contain the same quanta. For
example in a $B=4$ sector states \{C,1,2,2,5\} and \{D,1,2,5,2\}
belong to the same word while states \{C',4,3,3,8\} and
\{D',1,2,3,5\} are from different words. All gauge invariant
 classes, which can be constructed for particular set of spatial indices, are included
in a given word by definition. Hence a word is uniquely labelled
by an $ordered$ set of B integers, e.g. W=(1225) defines the word
considered above. How many states are in a word? As many as there
are inequivalent permutations of spatial indices which give
linearly independent states in all contributing gauge invariant
classes. Now Eq.(\ref{mix}) can be stated as the following
\newline THOREM 1: Only states in the same word mix.

Hence the whole mixing matrix splits into
blocks of small $(N_W \times N_W)$ mixing matrices for each word $W$ with $N_W$ being the
number
of states in a word $W$.

\noindent $Types$ \newline
Words come in different $types$. A type $T$ characterizes the pattern of indices
irrespectively of their actual value. For example words (1225) and (3448) belong
to the same type which we denote as [211], and (1234) or (2227) belong to the two
different types
[1111] and [31] respectively. The number of words in a given type, $N_T$ is equal to the
number of
realizations of the pattern by different values of the indices. For our first example
($d=D-1$)
\eq
N_{211} = d(d-1)(d-2)/2 ,\label{nwrds}
\eqx
and similarly for other types. Types can be identified with the partitions of $B$.
Therefore
the total number of types in a given bosonic sector is just equal to the
number of partitions,  $P_B$, of B.

Since neither the number of states in a word nor the number of
types in a sector depend  on the dimensionality of the problem,
the main complexity lies in the number of words in a given type,
e.g. Eq.(\ref{nwrds}) which grows rapidly with $d$ for larger $B$.
However the big simplification occurs due to the following
consequence of the Wick theorem. \newline THEOREM II: All mixing
matrices in the same type are equal. \newline To see this rewrite
$M_{IK}$ as the sum over all contractions \eq
M_{IK}=<I|K>=<0|a^{i_1}\dots a^{i_B} a^{\dagger}_{k_1}\dots
a^{\dagger}_{k_B}|0>= \Sigma_{contractions} \Pi \delta_{i.k.},
\eqx where the product on the right hand side denotes symbolically
given contraction. It is evident, even from this schematic
expression, that $M_{IK}$ does not depend on the particular
$values$ of the indices, but depends only on their
$configurations$ or patterns. Hence the theorem follows.

Therefore the calculation of the full mixing matrix is simply reduced to calculating
$P_B$ small matrices and using the same copy for all words in a given type.
Moreover, the tedious orthonormalizaton procedure also decouples into small blocks,
corresponding to words, and has to be done only for one word in a type with the rest being
the exact copies of the results from the first word.
This simplification prompts us to introduce the new scheme of organizing states in a basis
-- the TWS scheme which groups states according to the following hierarchy:
\eq
 S(tates)\rightarrow W(ords) \rightarrow T(ypes) \rightarrow Sectors .
\eqx With this organization any state in a basis is labelled by
four integers \eq |I>=|b,t,w,s>, \eqx with $b,t,w,s$ enumerating
sectors, types, words and states respectively.

It turns out that the structure of the hamiltonian is also much more transparent
in this representation as discussed below.

\subsection{SYMQM hamiltonian}

As above we shall consider only bosonic sectors $(B,F)=(B,0)$ where only $H_B$ of
Eq.(\ref{HD4})
contributes. Since $H_B$ contains terms of the second and fourth order in bosonic
creation and annihilation operators, it can only induce transitions with $\Delta B =0,2,4
$.
Each transition can be interpreted as the effective operator creating/annihilating 0,2 or
4
indices in our TWS representation. They are listed in Table 1 and will be referred to as
$ rules$ since each transition implies a rule how to generate from the initial type/word
the
final type/word with nonzero matrix element. For example $\Delta B=0$ transition, labelled
as rule 2,
annihilates two indices in the initial state and creates 2 indices with a common value
which
is new, i.e. does not exist in the initial state. Rule 3 has similar action but the common
value
of the created pair coincides with one of the values already existing in the initial
state.
Rule 7 adds two pairs of identical indices one being new and one already existing, etc.
It follows from Eq.(\ref{v}) that we cannot create/annihilate a pair of different indices.
Similarly there are no terms creating four the same indices.

  \begin{table}
  \begin{center}
   \begin{tabular}{ccc}
\hline\hline
   $\Delta B$ & rule & action    \\
   \hline\hline
   $ $ & $ 1 $ & diagonal \\
   $0$ & $ 2 $ & $2\rightarrow 2\; new$ \\
   $ $ & $ 3 $ & $2\rightarrow 2\; old$ \\
\hline
   $ $ & $ 4 $ & $add\; 2\; new $ \\
   $2$ & $ 5 $ & $add\; 2\; old $ \\
\hline
   $ $ & $ 6 $ & $add\; 2\times 2\; new-new $ \\
   $4$ & $ 7 $ & $add\; 2\times 2\; new-old $ \\
   $ $ & $ 8 $ & $add\; 2\times 2\; old-old $ \\
\hline\hline
   \end{tabular}
  \end{center}
\caption{Effective transitions induced by the bosonic hamiltonian
(\ref{HD4}) in nine space dimensions. } \label{tb1}
\end{table}

We emphasize that above classification should be used only to
identify all nonzero matrix elements of the hamiltonian. The
actual values of matrix elements will be calculated within the
original approach.

  \begin{table}  
\begin{center}  
\footnotesize
   \begin{tabular}{cc||c|cc|ccccc|ccccccccccc||}
     $b$ & $   $ & $ 1 $ & $    $ & $ 2 $  & $   $ & $   $ & $ 3 $ & $   $  & $   $ & $   $
& $    $ & $    $  & $    $ & $    $ & $ 4  $ & $    $  & $    $ & $    $  & $    $ & $
$   \\
   $ $ & $ t $ & $ 1 $ & $ 1  $ & $ 2 $  & $ 1 $ & $ 2 $ & $ 3 $ & $ 4 $  & $ 5 $ & $ 1 $
& $ 2  $ & $ 3  $  & $ 4  $ & $ 5  $ & $ 6  $ & $ 7  $  & $ 8  $ & $ 9  $  & $ 10 $ & $ 11
$   \\
   \hline\hline
   $1$ & $ 1 $ & $ 1 $ & $ 4  $ & $   $  & $   $ & $   $ & $ 6 $ & $   $  & $   $ & $   $
& $    $ & $    $  & $    $ & $    $ & $    $ & $    $  & $    $ & $    $  & $    $ & $
$   \\
\hline
   $ $ & $ 1 $ & $ 4 $ & $ 12 $ & $   $  & $ 5 $ & $   $ & $ 4 $ & $   $  & $   $ & $   $
& $    $ & $ 7  $  & $    $ & $    $ & $    $ & $ 6  $  & $    $ & $    $  & $    $ & $
$   \\
   $2$ & $ 2 $ & $   $ & $    $ & $ 1 $  & $   $ & $ 5 $ & $   $ & $ 4 $  & $   $ & $   $
& $    $ & $    $  & $ 8  $ & $    $ & $ 7  $ & $    $  & $    $ & $    $  & $    $ & $
$   \\
\hline
   $ $ & $ 1 $ & $   $ & $  5 $ & $   $  & $ 1 $ & $   $ & $ 2 $ & $   $  & $   $ & $ 5 $
& $    $ & $ 4  $  & $    $ & $    $ & $    $ & $    $  & $    $ & $    $  & $    $ & $
$   \\
   $ $ & $ 2 $ & $   $ & $    $ & $ 5 $  & $   $ & $ 13$ & $   $ & $ 2 $  & $   $ & $   $
& $ 5  $ & $    $  & $ 5  $ & $    $ & $  4 $ & $    $  & $    $ & $    $  & $    $ & $
$   \\
   $3$ & $ 3 $ & $ 6 $ & $  4 $ & $   $  & $ 2 $ & $   $ & $ 12$ & $   $  & $   $ & $   $
& $    $ & $ 5  $  & $    $ & $    $ & $    $ & $  $4$ $  & $    $ & $    $  & $    $ & $
$   \\
   $ $ & $ 4 $ & $   $ & $    $ & $ 4 $  & $   $ & $ 2 $ & $   $ & $ 12$  & $   $ & $   $
& $    $ & $    $  & $    $ & $ 5  $ & $  5 $ & $    $  & $    $ & $ 4  $  & $    $ & $
$   \\
   $ $ & $ 5 $ & $   $ & $    $ & $   $  & $   $ & $   $ & $   $ & $   $  & $ 1 $ & $   $
& $    $ & $    $  & $    $ & $    $ & $    $ & $    $  & $  5 $ & $    $  & $ 4  $ & $
$   \\
\hline
   $ $ & $ 1 $ & $   $ & $    $ & $   $  & $ 5 $ & $   $ & $   $ & $   $  & $   $ & $ 1 $
& $    $ & $ 2  $  & $    $ & $    $ & $    $ & $    $  & $    $ & $    $  & $    $ & $  $
  \\
   $ $ & $ 2 $ & $   $ & $    $ & $   $  & $   $ & $ 5 $ & $   $ & $   $  & $   $ & $   $
& $ 1  $ & $    $  & $ 3  $ & $    $ & $  2 $ & $    $  & $    $ & $    $  & $    $ & $  $
  \\
   $ $ & $ 3 $ & $   $ & $  7 $ & $   $  & $ 4 $ & $   $ & $ 5 $ & $   $  & $   $ & $ 2 $
& $    $ & $ 123$  & $    $ & $    $ & $    $ & $  2 $  & $    $ & $    $  & $    $ & $  $
 \\
   $ $ & $ 4 $ & $   $ & $    $ & $ 8 $  & $   $ & $ 5 $ & $   $ & $   $  & $   $ & $   $
& $ 3  $ & $    $  & $ 1  $ & $    $ & $  2 $ & $    $  & $    $ & $    $  & $    $ & $  $
  \\
   $ $ & $ 5 $ & $   $ & $    $ & $   $  & $   $ & $   $ & $   $ & $ 5 $  & $   $ & $   $
& $    $ & $    $  & $    $ & $ 1  $ & $  3 $ & $    $  & $    $ & $ 2  $  & $    $ & $  $
  \\
   $4$ & $ 6 $ & $   $ & $    $ & $ 7 $  & $   $ & $ 4 $ & $   $ & $ 5 $  & $   $ & $   $
& $ 2  $ & $    $  & $ 2  $ & $ 3  $ & $ 123$ & $    $  & $    $ & $ 2  $  & $    $ & $  $
  \\
   $ $ & $ 7 $ & $   $ & $  6 $ & $   $  & $   $ & $   $ & $ 4 $ & $   $  & $   $ & $   $
& $    $ & $ 2  $  & $    $ & $    $ & $    $ & $ 12 $  & $    $ & $    $  & $    $ & $  $
  \\
   $ $ & $ 8 $ & $   $ & $    $ & $   $  & $   $ & $   $ & $   $ & $   $  & $ 5 $ & $   $
& $    $ & $    $  & $    $ & $    $ & $    $ & $    $  & $ 13 $ & $    $  & $ 2  $ & $  $
  \\
   $ $ & $ 9 $ & $   $ & $    $ & $   $  & $   $ & $   $ & $   $ & $ 4 $  & $   $ & $   $
& $    $ & $    $  & $    $ & $ 2  $ & $  2 $ & $    $  & $    $ & $ 12 $  & $    $ & $  $
  \\
   $ $ & $10 $ & $   $ & $    $ & $   $  & $   $ & $   $ & $   $ & $   $  & $ 4 $ & $   $
& $    $ & $    $  & $    $ & $    $ & $    $ & $    $  & $  2 $ & $    $  & $ 12 $ & $  $
  \\
   $ $ & $11 $ & $   $ & $    $ & $   $  & $   $ & $   $ & $   $ & $   $  & $   $ & $   $
& $    $ & $    $  & $    $ & $    $ & $    $ & $    $  & $    $ & $    $  & $    $ & $ 1
$   \\
\hline\hline
   \end{tabular}   
\caption{Structure of the bosonic hamiltonian in the TWS representation,
$b$ and $t$ are defined in Table 3. } \label{tab2}
  \end{center}
\end{table}

  \begin{table}
  \begin{center}
   \begin{tabular}{||cc|cc||}
\hline\hline
   $b$ & $ t $ & $ B $ & $ partition $    \\
   \hline\hline
   $1$ & $ 1 $ & $ 0 $ & $   $ \\
\hline
   $ $ & $ 1 $ & $   $ & $ 2 $ \\
   $2$ & $ 2 $ & $ 2 $ & $ 1 $ \\
\hline
   $ $ & $ 1 $ & $   $ & $ 4 $ \\
   $ $ & $ 2 $ & $   $ & $31 $ \\
   $3$ & $ 3 $ & $ 4 $ & $22 $ \\
   $ $ & $ 4 $ & $   $ & $211 $ \\
   $ $ & $ 5 $ & $   $ & $1111$ \\
\hline
   $ $ & $ 1 $ & $   $ & $ 6 $\\
   $ $ & $ 2 $ & $   $ & $ 51 $\\
   $ $ & $ 3 $ & $   $ & $ 42$\\
   $ $ & $ 4 $ & $   $ & $ 33$\\
   $ $ & $ 5 $ & $   $ & $411$\\
   $4$ & $ 6 $ & $ 6 $ & $321$\\
   $ $ & $ 7 $ & $   $ & $222$\\
   $ $ & $ 8 $ & $   $ & $3111$\\
   $ $ & $ 9 $ & $   $ & $2211$\\
   $ $ & $10 $ & $   $ & $21111$\\
   $ $ & $11 $ & $   $ & $111111$\\
\hline\hline
   \end{tabular}
  \end{center}
\caption{Labelling of sectors and types in Table 2. } \label{tab3}
\end{table}

Table 2 shows the resulting structure of the hamiltonian in the
TWS representation for the first four sectors. Labelling of rows
and columns is explained in Table 3. The hamiltonian is block
tri-diagonal in bosonic sectors (odd and even sectors are
decoupled bacuse of the parity conservation). Each nontrivial
sector-to-sector block is again sparse with only few nonzero
subblocks connecting initial and final types according to the
rules listed in Table 1. For simplicity they are labelled by these
rules. Each such type-to-type subblock is again a $N_{Tf}\times
N_{Ti}$ matrix with entries connecting different words of the
initial and final types. Again this matrix is in general sparse -
from given initial word one can reach, with a given rule, only few
final words. This is not the end, the entries of this matrix are
again matrices labelled by the individual states belonging to
initial and final words. Only at this level the orthogonalization
of basis states enters. Since elementary states in a single word
are not orthogonal, the word-to-word blocks must be corrected by
the square roots of the corresponding norm matrices. This is,
however, much less time consuming than the global
orthonormalization.

Moreover one can show that the word-to-word subblocks in a given type-to-type block are
independent of the initial word. Similarly to the mixing matrix case, it suffices to
calculate
transitions from one initial word in each type and substitute them for other initial
words.
The proof is more technical and will not be given here.

The whole structure, even though little involved at first glance,
can be readily formalized and leads to much faster calculations of the whole matrix.

\subsection{Preliminary results}

The complexity of the problem is summarized in Table 4 where dimensions of the Hilbert
space
for various $D$ and $B$ are displayed. All bases were generated and orthogonalized
with the original approach (c.f. Section 4.1). Mathematica programs took about 2 days to
produce the last entry, thereby the cutoff $B = 7 - 8$ seems to be a practical maximum for
the brute force method at $D=10$. Calculation and diagonalization of the hamiltonian is
yet
more demanding - for D=10 one can reach only $B=4$ within the reasonable computing time
\cite{JW2}.

  \begin{table}
  \begin{center}
   \begin{tabular}{cccccccc} \hline\hline
        $D$        &  $ 4 $ & $ 5 $ & $ 6 $ & $ 7 $  & $ 8 $   & $ 9 $ & $ 10 $  \\  \hline
  $B$ & $   $ & $   $ & $   $ & $ N_{s}$ \ \ & $  $ & \ \ $   $ & \ \ $    $   \\
   \hline
  0 &     1\ \ &    1    &  1      &     1          &    1       &     1          &  1
\\
  1 &     - \ \ &    -  &  -      &     -         &    -        &     -         &  -   \\
  2 &     6 \ \ &   10  &   15   &    21        &   28       &     36       &  45   \\
  3 &     1 \ \ &    4   &   10   &    20         &   35      &     56       &  84     \\
  4 &    21 \ \ &   55 &  120   &   231      &  406      &    666     &  1035 \\
  5 &     6 \ \ &   36   &  126  &   336       &  756      &   1512    &  2772 \\
  6 &    56 \ \ &  220  &  680  &  1771     & 4060    &   8436   & 16215 \\
  7 &    21 \ \ &  180   &  855  &  2976    & 8478   &  20952   &             \\
  8 &   126 \ \ &  714  & 3045  & 10521  &            &                 &              \\
   \hline\hline
   \end{tabular}
   \end{center}
\caption{Sizes of bases in the $F=0$ sector for space-time dimensions $4\le D \le 10$.
$N_s$ is the
number of basis vectors
  with given number of bosonic quanta, $B$. }
\label{tab4}
\end{table}

The TWS approach allows to reach one to two orders of magnitude bigger bases, which
at D=10 translates for $B_{max} = 6 - 8$. This rather modest, in terms of $B$, improvement
is nevertheless quite relevant for the precision of the energy levels.
Fig. 4 shows the first five eigenenergies as a function of the cutoff.
Solid lines are drawn just to guide the eye. Levels are labelled by corresponding SO(9)
representations.
Since our cutoff respects the rotational invariance, resulting spectrum of a cut
hamiltonian should have full SO(9) symmetry \footnote{Only in the purely bosonic sector.},
and indeed we observe first few SO(9) multiplets as indicated in the Figure
\footnote{Incidentally this provides one of the tests of the whole TWS scheme.}.
The ordering of the levels (singlet, tensor, singlet) turns out to be the same
as in the well known $D=4$ case \cite{LM} and in fact remains unchanged for all
$4\le D \le 10$ \cite{JW2} .
The energy of the lowest singlet state has already converged within 3\%.
For second and third state
the relative change with $B$ dropped from 13\% to 7\% and from 17\% to 10\% respectively.
Cutoff $B_{max}=8$ can be reached with the reasonable computing effort.
\begin{figure}[htb]
\epsfig{width=12cm,file=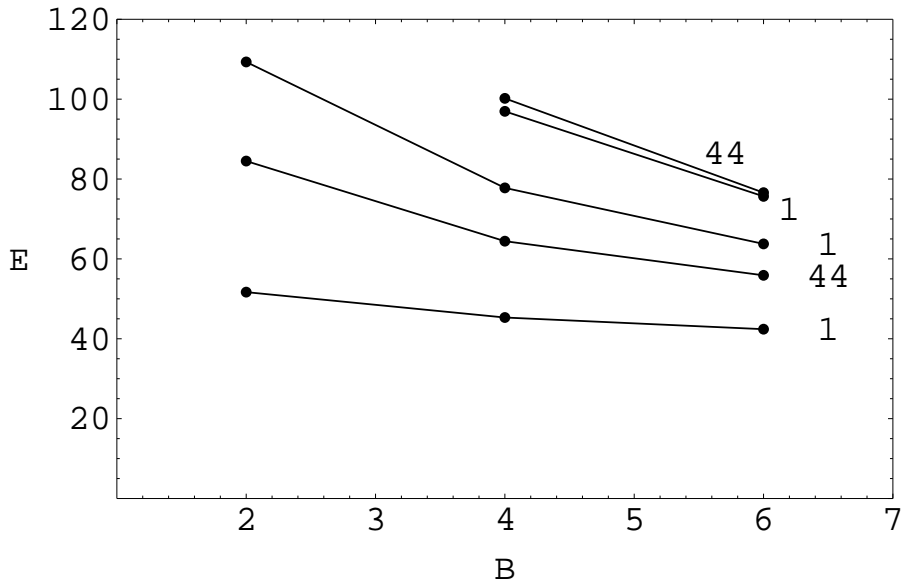}
\caption{ Lower levels of the spectrum of the pure Yang-Mills D=10
system.} \label{f4}
\end{figure}

\section{Summary and outlook}

In the last year there has been a substantial progress in solving supersymmetric
Yang-Mills quantum mechanics with the SU(2) gauge group. In four dimensions
we can reach such high cutoffs
that the restriction of the Hilbert space becomes irrelevant for most purposes.
Precise spectrum in all channels has been obtained and the intricate pattern
of continuous and discrete states found and clarified. Supersymmetric structure
of the system can be now studied in detail. In particular the complete classification
of supersymmetric multiplets is now available.

In the ten dimensional system developments are obviously slower
but neverheless steady. In the bosonic sector one can now reach
the cutoffs (hence the precision) comparable to the $D=4$ model
studied with the original technique. The new TSW scheme is yet to
be generalized to fermionic sectors. The ongoing puzzle of
nonconserved massless Majorana fermions calls for further study.
On the other hand reported here progress in the four dimensional
case opens some new possibilities also for the ten dimensional
system. \vspace*{.3cm}

\noindent {\em Acknowledgments}
I thank the Theory Group of the Max-Planck-Institut in Munchen for their hospitality.
This work is  supported by the
Polish Committee for Scientific Research under the grant no. PB 2P03B09622,
during 2002 - 2004.

\end{document}